\documentclass[a4paper,11pt]{article}
\usepackage{pos}

\title{Chaos, Cosmic Ray Anisotropy, and the Heliosphere}
\author*[a]{Vanessa L\'opez-Barquero}
\author[b]{Paolo Desiati}

\affiliation[a]{Institute of Astronomy, University of Cambridge,\\
 Madingley Road, Cambridge CB3 OHA, UK}

\affiliation[b]{Wisconsin IceCube Particle Astrophysics Center (WIPAC), University of Wisconsin-Madison\\
Madison, WI 53703, USA}

\emailAdd{v.lopezbarquero@ast.cam.ac.uk}
\emailAdd{desiati@icecube.wisc.edu}

\abstract{After more than a century of discovering cosmic rays, a comprehensive description of their origin, propagation, and composition still eludes us. One of the difficulties is that these particles interact with magnetic fields; therefore, their directional information is distorted as they travel. In addition, as cosmic rays (CRs) propagate in the Galaxy, they can be affected by magnetic structures that temporarily trap them and cause their trajectories to display chaotic behavior, therefore modifying the simple diffusion scenario.

Here, we examine the effects of chaos and trapping on the TeV CR anisotropy. Concretely, we develop a new method to study the chaotic behavior of CRs. This work is based on the heliospheric effects since they can be remarkably significant for this anisotropy. Specifically, how the distinct heliospheric structures can affect chaos levels. We model the heliosphere as a coherent magnetic structure given by a static magnetic bottle and the presence of temporal magnetic perturbations. This configuration is used to describe the draping of the local interstellar magnetic field lines around the heliosphere and the effects of magnetic field reversals induced by the solar cycles.

In this work, we explore the possibility that particle trajectories may develop chaotic behavior while traversing and being temporarily trapped in this heliospheric-inspired toy model and the potential consequences it can have on the cosmic ray arrival distribution. It was found that the level of chaos in a trajectory is linked to the time the particles remain trapped in the system. This relation is described by a power law that could prove to be inherently characteristic of the system. Also, the arrival distribution maps show areas where the different chaotic behaviors are present, which can constitute a source of time-variability in the CR maps and can prove critical in understanding the anisotropy on Earth.}

\FullConference{%
  *** 27th European Cosmic Ray Symposium - ECRS ***\\
  *** 25-29 July 2022 ***\\
  *** Nijmegen, the Netherlands ***
}

\begin{document}
\maketitle

\section{Introduction}

Galactic cosmic rays are detected on Earth with anisotropy in their arrival directions~\citep{2016ApJ...826..220A, Abeysekara_2019}. Multiple experiments have measured this anisotropy; however, a complete explanation remains elusive. This work investigates the effects of chaotic trajectories of trapped cosmic rays on this arrival anisotropy. Specifically, how a coherent structure, such as the heliosphere, can influence particle propagation and, ultimately, the directions in which particles are detected.

\section{Magnetic Field Configuration}
To simulate the trapping conditions in a magnetic field, we developed a model consisting of an axial-symmetric magnetic bottle with magnetic time perturbations. The temporal perturbations to the magnetic field have the following form:

\begin{equation}
B_{y} = \frac{\Delta B}{B}\,\sin(k_p x-\omega_p t)\,e^{-\frac{1}{2}\left(\frac{z}{\sigma_p}\right)^2},
\label{eq:pert}
\end{equation}
where $k_p = \frac{2\pi}{L_p}$ and $\omega = \frac{2\pi v_p}{L_p}$.

The model's specific parameters, such as the magnetic field strength and velocities, are comparable to the heliospheric values. The magnetic bottle is based on the mirroring effect that particles experience when they bounce off the heliosphere's flanks. The time perturbations mimic the effects of magnetic field reversals caused by the 11-year solar cycles.

We created three different systems to test how the different components can affect the particles' trajectories. One is the \textit{unperturbed system}, which consists of just the magnetic bottle. With this system, we will assess the effects of mirroring and trapping. The second is the \textit{weak perturbation system} which corresponds to the magnetic bottle plus the time perturbation to the magnetic field, with the parameters chosen as those expected for the heliosphere: $\frac{\Delta B}{B} = 0.1$ and $v_p = 2$ AU/year. Finally, a third one, the \textit{strong perturbation system}, is created by increasing the values of the weak perturbation in order to amplify the effects so that they will be easy to distinguish for the analysis: $\frac{\Delta B}{B} = 0.5$ and $v_p = 20$ AU/year.

\section{Chaos and the Finite-Time Lyapunov Exponent}
We develop a new method for characterizing chaos in the trapping conditions of this magnetic structure. With this new method, we can assess the chaotic effects on particle trajectories and how they are affected by being temporarily trapped in coherent structures.

We based our model on the Finite-Time Lyapunov Exponent (FTLE): 

\begin{equation}\label{FTLE}
\lambda (t,\Delta t)=   \frac{1}{\Delta t}   \ln \left [ \frac{d(t+\Delta t)}{d(t) } \right ],
\end{equation}
where $\Delta t$ is the time interval for the calculation and $d$ is the distance in phase space between two particles at a specific time.

As a result, the FTLE can calculate the level of chaos based on the divergence rate of the trajectories. This quantity is useful because it can adapt to temporarily trapped conditions caused by interactions with coherent magnetic structures.

\section{Results and Discussion}
Once particles are propagated in this system, according to the method described in~\cite{barquero_2019}, we found a relation between the Finite-time Lyapunov exponent (FTLE), which represents the chaotic behaviour of the particles, and the time required to escape the system. A power law provides this correlation:

\begin{equation}
\lambda_{\tiny{FTLE}} = \beta\, t_{esc}^{-1.04 \pm 0.03} .
\label{eq:powerlaw}
\end{equation}

One remarkable property of these systems is that the same power law persists even when perturbations are introduced. This feature points to the idea that a system can be characterized by a particular law, which will help in our aim to understand the role of trapping and chaotic behavior in cosmic ray propagation.

\begin{figure*}
	\centering
	\includegraphics[width=1\linewidth]{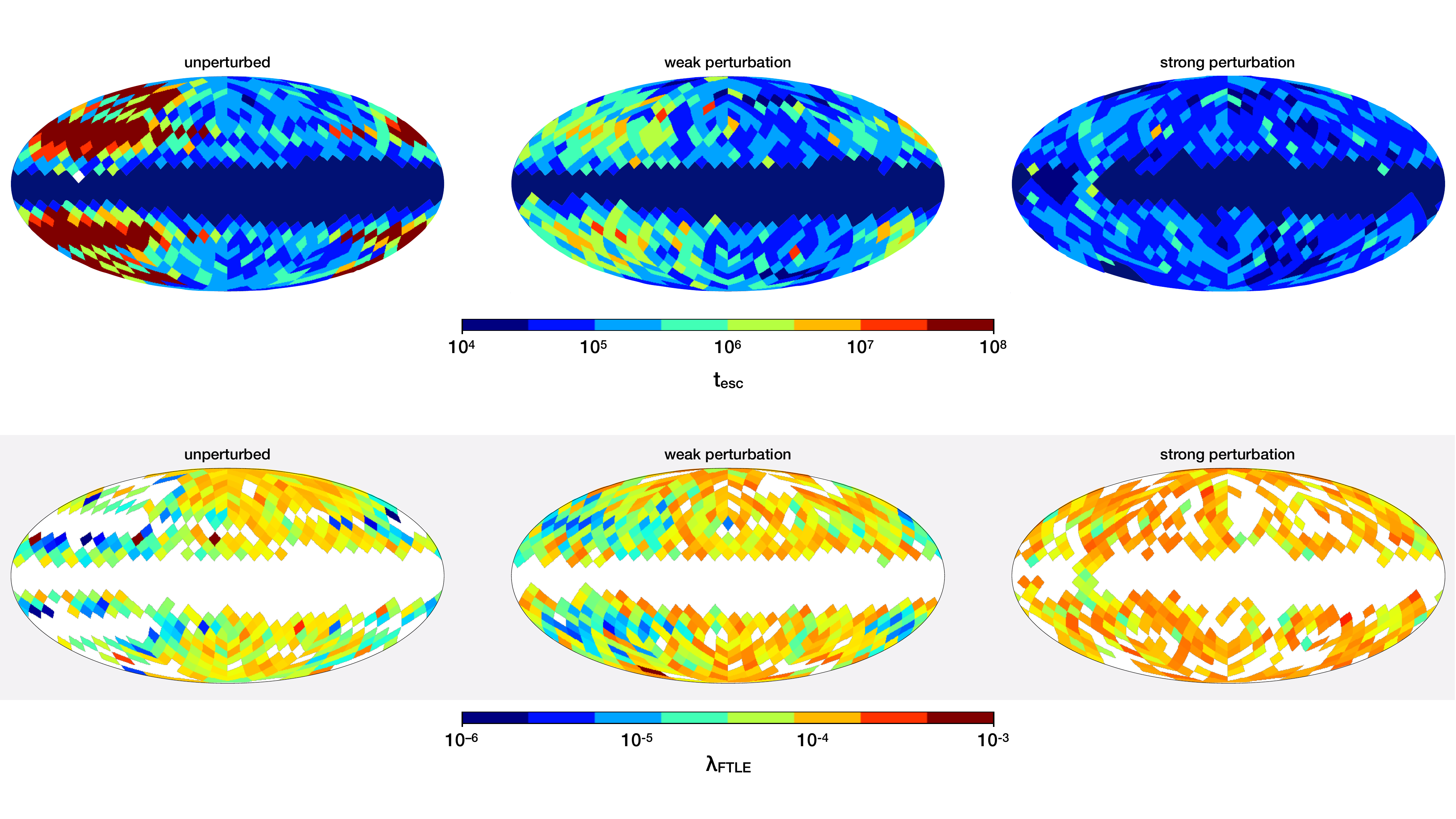}
	\caption{ \textit{Maps.} Each pixel in a map corresponds to a particle's arrival direction, and the values for each pixel represent the escape time or finite-time Lyapunov exponent. The escape times for the unperturbed, weakly perturbed, and strongly perturbed systems are presented in the top panel, respectively. The bottom panel corresponds to the finite-time Lyapunov exponent, $\lambda_{\tiny{FTLE}}$. The white pixels in the bottom panel are for non-chaotic particles.}
	\label{fig:figure-skymaps}
\end{figure*}

The Finite-Time Lyapunov exponents and escape times are plotted in arrival distribution maps to derive information that will help us interpret the observations (see Figure~\ref{fig:figure-skymaps}). These maps constitute a visual representation of the various chaotic behaviors and how they are spatially distributed. For example, we can see areas of the unperturbed map where particles are more chaotic (denoted in redder colors in the bottom panel) and less chaotic areas near the stability region (darker blue). 

As the time perturbation strengthens (left to right in the maps), we can see how the maps change accordingly. The more chaotic particles start to populate larger regions of the map. In the case of the heliosphere, we can expect maps similar to those in the middle panel. There will be a mix of the level of chaos for the particles in the weak perturbation, and there will be regions with more significant variations due to the chaotic nature of the particles in it.

These maps show areas with various chaotic behaviours, which may affect the observations. For instance, this might be a source of temporal variability. Since the areas where the particles are very chaotic will change differently than the ones where non-chaotic particles reside.

\end{document}